\documentclass[10pt,a4paper,hidelinks]{article}

\usepackage{fancyhdr}
\usepackage{pdfpages}

\fancypagestyle{CISPA}{

\fancyhead[C]{\includegraphics[width=4cm]{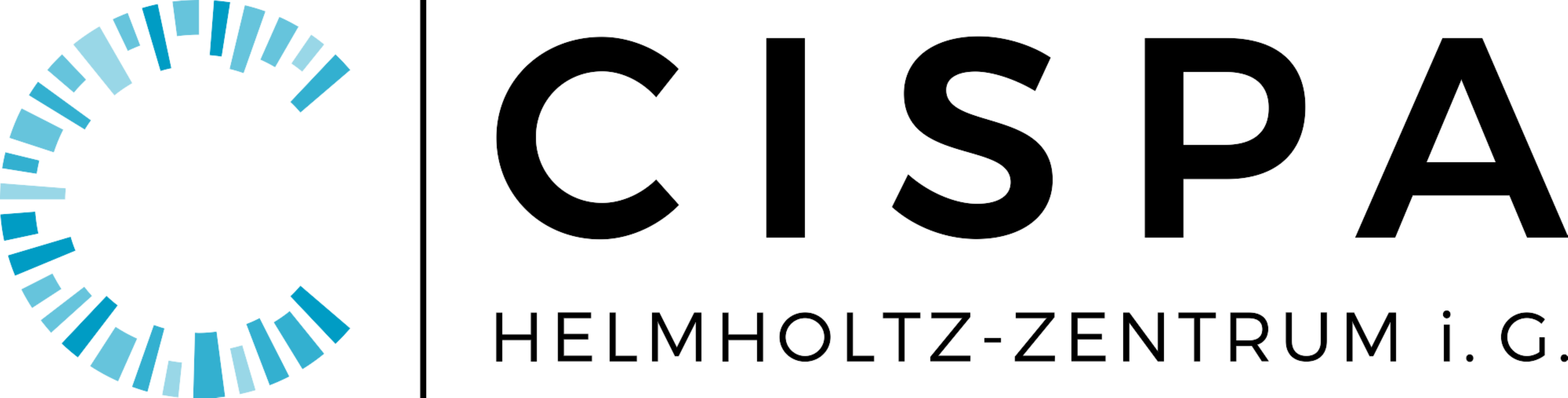}}
}
\usepackage{csquotes}
\usepackage[backend=biber]{biblatex}
\usepackage{etoolbox}

\usepackage[T1]{fontenc}
\usepackage{inputenc}
\usepackage[english]{babel}
\usepackage[nohead,includefoot,margin=2cm]{geometry}
\usepackage[compact,raggedright,small]{titlesec}
\usepackage[affil-it]{authblk}
\usepackage[runin]{abstract}
\usepackage{multicol}
\usepackage{titling}
\usepackage{graphicx}
\usepackage{montserrat}
\usepackage{setspace}
\usepackage{booktabs}   %
\usepackage{multirow}
\usepackage{subcaption} %
\usepackage{caption}
\usepackage{tikz}
\usetikzlibrary{arrows}
\usetikzlibrary{positioning}

\usepackage{xspace}
\usepackage{hyperref}  %
\usepackage[nameinlink,noabbrev]{cleveref}  %
\usepackage[inline]{enumitem}
\usepackage{microtype} %
\usepackage{alltt}
\usepackage{float}
\usepackage{stfloats}

\usepackage[noend]{algpseudocode}
\usepackage{listings}
\usepackage{color}
\usepackage{relsize} %

\definecolor{codegreen}{rgb}{0,0.6,0}
\definecolor{codegray}{rgb}{0.5,0.5,0.5}
\definecolor{codepurple}{rgb}{0.58,0,0.82}
\definecolor{backcolour}{rgb}{0.95,0.95,0.92}

\lstdefinestyle{mystyle}{
    commentstyle=\color{codegreen},
    keywordstyle=\color{magenta},
    numberstyle=\tiny\color{codegray},
    stringstyle=\color{codepurple},
    basicstyle=\footnotesize,
    breakatwhitespace=false,
    breaklines=true,
    captionpos=b,
    keepspaces=true,
    numbers=left,
    numbersep=5pt,
    showspaces=false,
    showstringspaces=false,
    showtabs=false,
    tabsize=2
}

\algblockdefx[switch]{Switch}{EndSwitch}%
[1]{\textbf{switch} #1}%
{}%
\algloopdefx{Case}[1]{\textbf{case} #1:}%
\algdef{SE}[DOWHILE]{Do}{doWhile}{\algorithmicdo}[1]{\algorithmicwhile\ #1}%

\def\term#1{\texttt{"\textbf{#1}"}}
\def\nonterm#1{\textnormal{\emph{#1}}}
\def\regex#1{\texttt{{\color{blue}/}#1{\color{blue}/}}}
\def\expandsto{\(\rightarrow{}\)}

\def\|#1|{\textit{#1}}
\def\<#1>{\texttt{#1}}

\newcommand{\todo}[1]{\textcolor{red}{\textbf{TODO }\textit{#1}}}
\newcommand{\done}[2]{[#1]\textcolor{blue}{\textbf{DONE}~\textit{#2}}}

\renewcommand{\todo}[1]{}
\renewcommand{\done}[2]{}

\newcommand{\pyg}{\textsc{Pygmalion}\xspace}
\newcommand{\afl}{\textsc{afl}\xspace}
\newcommand{\klee}{\textsc{Klee}\xspace}
\newcommand{\pdtg}{parser-directed test generation\xspace}
\newcommand{\json}{{\smaller JSON}\xspace}
\newcommand{\SC}{Statement Coverage\xspace}

\newcommand{\ul}{{\smaller URL}\xspace}
\newcommand{\me}{mathematical expressions\xspace}
\newcommand{\mes}{\textit{Mathexpr}\xspace}
\newcommand{\microjsonpy}{\textit{microjson.py}\xspace}

\title{Sample-Free Learning of Input Grammars for~Comprehensive~Software~Fuzzing}    %
\date{\small (Dated \today)}
\author{Rahul Gopinath}
\author{Bj\"orn Mathis}
\author{Mathias H\"oschele}
\author{Alexander Kampmann}
\author{Andreas Zeller}
\affil{\{rahul.gopinath, bjoern.mathis, hoeschele, kampmann, zeller\}@cispa.saarland \\
CISPA / Saarland University, Saarland Informatics Campus, Saarbr\"ucken, Germany}

\newcommand\BackgroundPic{
    \put(0,0){
    \parbox[b][\paperheight]{\paperwidth}{%
    \vfill
    \centering
    \includegraphics[width=\paperwidth,height=\paperheight]{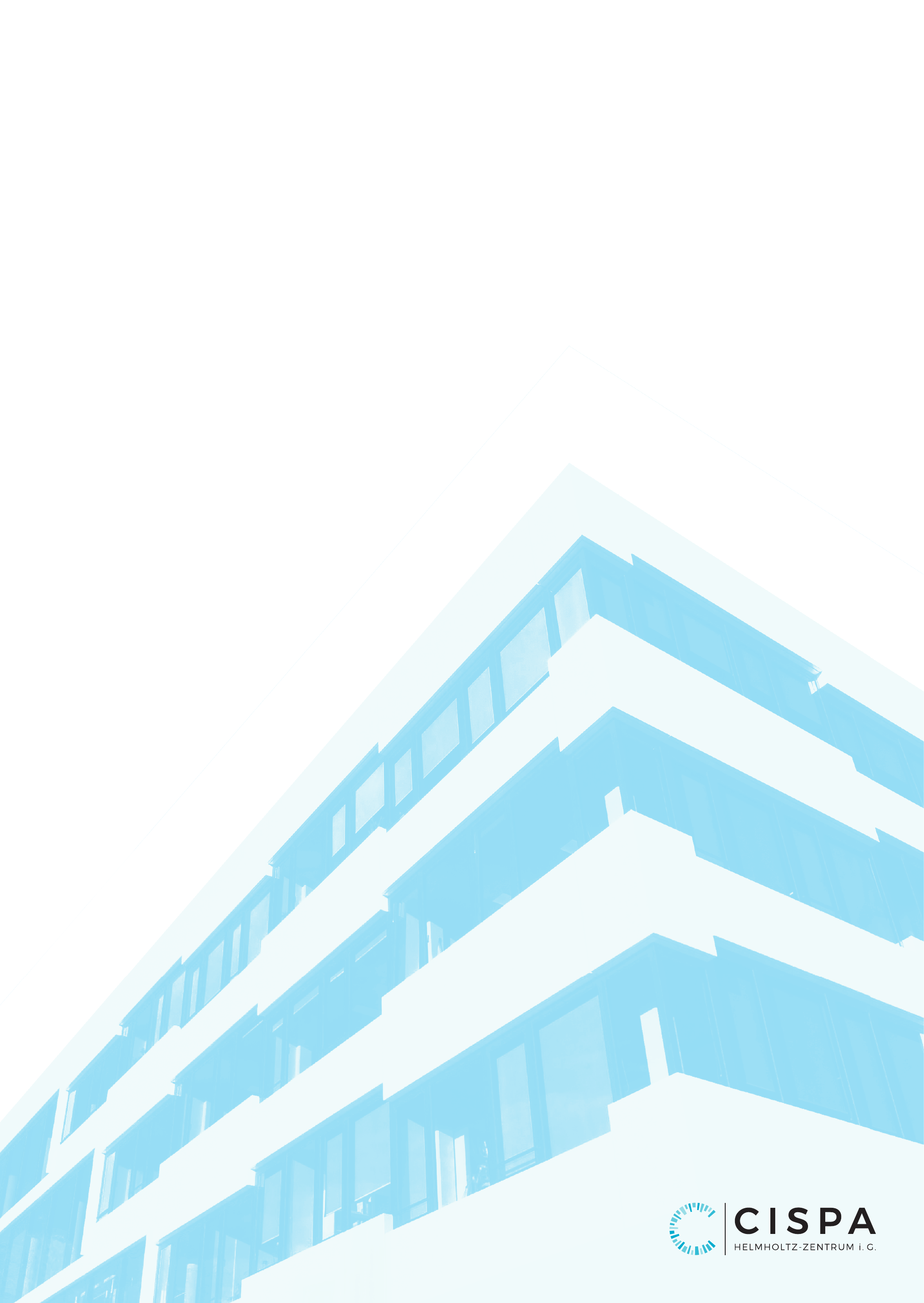}
    \vfill
}}}

\usepackage{utopia}
\usepackage{times}

\begin{document}
\AddToShipoutPicture*{\BackgroundPic}

\makeatletter
\renewcommand{\Authfont}{\normalsize\sffamily\bfseries}
\renewcommand{\Affilfont}{\normalsize\sffamily\mdseries}
\begin{titlepage}
\newcommand{\HRule}{\rule{\linewidth}{0.1mm}}
\centering
  \textsc{\LARGE {\fontfamily{Montserrat-TOsF}\selectfont CISPA Helmholtz-Zentrum i.G.}}\\[1.5cm]

  \vspace{2.4 cm}
  \HRule \\[0.2cm]
  {\huge\sffamily\bfseries \@title\par}
  \vspace{0.2cm}
  \HRule \\[1.5cm]

  {\sffamily \@author\par}
\vfill

\end{titlepage}

\makeatother
\setlength{\affilsep}{0.1em}
\addto{\Affilfont}{\small}
\renewcommand{\Authfont}{\normalsize}
\renewcommand{\Affilfont}{\normalsize}

\pretitle{\begin{center}\large\bfseries}
\posttitle{\end{center}}
\maketitle
\thispagestyle{CISPA}

\begin{abstract}
Generating valid test inputs for a program is much easier if one knows the input language. We present first successes for a technique that, given a program~$P$ without any input samples or models, learns an \emph{input grammar} that represents the syntactically valid inputs for~$P$---a grammar which can then be used for highly effective test generation for~$P$. To this end, we introduce a \emph{test generator targeted at input parsers} that systematically explores parsing alternatives based on dynamic tracking of constraints; the resulting inputs go into a \emph{grammar learner} producing a grammar that can then be used for fuzzing.
In our evaluation on subjects such as \json, \ul, or \mes, our \pyg prototype took only a few minutes to infer grammars and generate thousands of valid high-quality inputs.
\end{abstract}
\begin{multicols}{2}

\section{Introduction}

Testing programs with generated inputs is a common way to test programs for robustness.  Such generated inputs must be \emph{valid,} because otherwise, they would be rejected by the program under test before reaching the functionality to be tested; and they must well sample \emph{the full range of possible inputs,} because otherwise, important program features may not be covered.  In the absence of a formal input specification such as a grammar, common test generators have to rely on \emph{samples of valid inputs.}  These would then
\begin{enumerate*}
\item be systematically mutated~\cite{AFLFuzz} using generic operations such as bit flips or character exchanges; or 
\item be used to \emph{infer grammars and syntactical rules} that can then be used to generate more similar inputs~\cite{hoschele2016mining, bastani2017synthesizing, godefroid2017learnandfuzz}.
\end{enumerate*}
Both approaches, however, would have great difficulty synthesizing features that are not present in the original samples already.  In principle, test generators could use \emph{symbolic analysis} on the program under test to determine and solve the exact conditions under which an input is accepted~\cite{visser2004test, sen2005cute, cadar2008klee, chen2018angora}; but nontrivial input formats induce a large number of constraints that can easily overwhelm symbolic constraint solvers.

\begin{figure*}

\tikzset{
    ultra thin/.style= {line width=0.1pt},
    very thin/.style=  {line width=0.2pt},
    thin/.style=       {line width=0.4pt},%
    semithick/.style=  {line width=0.6pt},
    thick/.style=      {line width=0.8pt},
    very thick/.style= {line width=1.2pt},
    ultra thick/.style={line width=2pt}
}
\begin{tikzpicture}[->,>=stealth',shorten >=1pt,auto,node distance=3cm,
  thick,main node/.style={font=\sffamily\bfseries,minimum size=15mm},label node/.style={font=\sffamily\bfseries,minimum size=5mm,draw,circle,gray!80}]

  \node[main node] (PUT) {Program Under Test};
  \node[main node] (PDTG) [below of=PUT] {Parser-Directed Test Generator};
  \node[main node] (GL) [right=20em of PDTG] {Grammar Learner};
  \node[main node] (F) [right=25em of PUT] {Fuzzer};
  
  \node[label node] (0) [left=0.1em of PUT] {0};
  \node[label node] (1) [left=0.1em of PDTG] {1};
  \node[label node] (2) [right=0.1em of GL] {2};
  \node[label node] (3) [right=0.1em of F] {3};
  
  \path[every node/.style={font=\sffamily,
  		fill=white,inner sep=1pt},
        every edge/.style={draw=red,ultra thick}]
    (PDTG) edge node [anchor=center,pos=0.5] {Inputs + Equivalence Classses} (GL)
    (PUT)  edge [left=55] node[right=1em] {Comparisons} (PDTG)
    (PUT)  edge [bend right=5] node[anchor=center,pos=0.5] {Dynamic Taints} (GL)

    (GL)   edge [bend right=30] node[left=1em] {Grammar} (F)
    (F)    edge [ right=40] node[left=1em] {Test Inputs} (PUT)
    (PDTG) edge [bend left=50] node[left=1em] {Valid Prefixes} (PUT);
\end{tikzpicture}
\caption{The \pyg prototype starts with a program under test~(0) into which we feed a fixed, valid prefix (say, an empty string).  By dynamically tracking the comparisons of input characters against expectations, a \emph{parser-directed test generator}~(1) systematically satisfies these expectations, eventually producing a set of inputs that cover all parser alternatives.  These go into a \emph{grammar learner}~(2), which by tracking the data flow of these characters through the program produces an input grammar.  Using this grammar, a \emph{fuzzer}~(3) can now produce syntactically valid program inputs at high speed, systematically covering input features.}
\label{fig:overview}
\end{figure*}
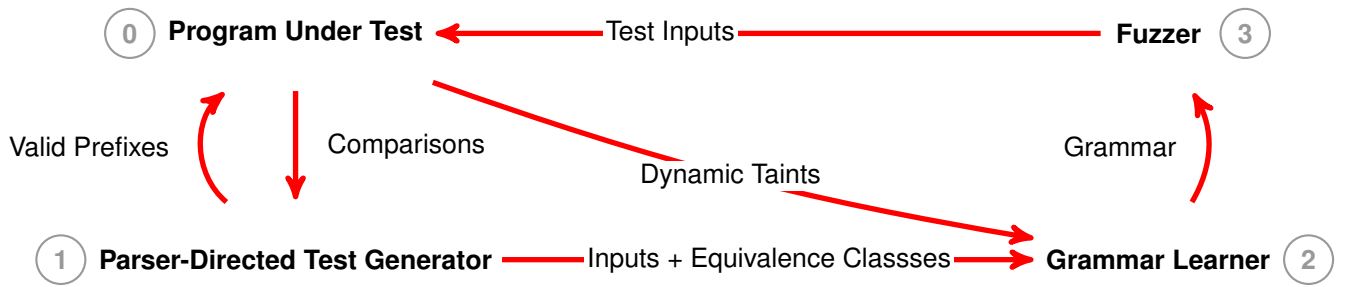

In this paper, we follow recent advances in grammar inference~\cite{hoschele2016mining, bastani2017synthesizing, godefroid2017learnandfuzz} by first learning an input grammar, and then using this grammar for test generation.  In contrast to this state of the art, however, our approach automatically infers an accurate description of the input language \emph{without requiring any input samples at all---}actually, all that is needed for comprehensive testing is the program itself.  \Cref{fig:overview} summarizes our approach:

\begin{enumerate}
  \item To address the problem of learning without samples, we introduce a \emph{test generator specifically targeting input parsers.}  Our approach starts with a fixed input (typically an empty string), which would be rejected.  During parsing, we use dynamic tainting\footnote{
We use a \emph{pure python} library similar to the algorithm by Conti et al.~\cite{conti2010a} for tracing taints and comparisons. Hence, our algorithm \emph{does not} require specially crafted interpreter to track taints.} to dynamically track all comparisons of input characters against expected values, and then provide an input that satisfies these expectations.  By repeating this process from rejection to rejection, we eventually obtain a set of inputs that covers all comparison alternatives made by the parser---and consequently, all structural (syntactic) alternatives as well.  %

\item To \emph{learn a grammar} from the parser-covering inputs, we  dynamically track the \emph{data flow of input characters} throughout program execution to induce a grammar.  Our main algorithm is inspired by Hoschele et al.~\cite{hoschele2016mining}: Character sequences that share the same data flow then form syntactic entities; subsequences with different data flow induce composition rules.  On top, our grammar learner makes use of \emph{equivalence classes} found during Step~1: If the test generator finds that, say, some input fragment can be any digit, this generalization is also reflected in the grammar.

\item To \emph{produce inputs,} we use the grammar from Step~2 as a \emph{producer,} now very rapidly producing inputs for the program under test.  At this point, no instrumentation of the program is required anymore, and the inputs produced could also be given to another program with the same input language.
\end{enumerate}

As a result, we obtain a tool chain that \emph{requires nothing but an executable code, and produces high-quality inputs that cover and combine all syntactic features.}  Our \pyg prototype\footnote{PYGMALION = PYthon Grammar Miner 
Actively Learning Inputs Of Note.  See also Pygmalion~\cite{shaw1916pygmalion} on language learning..
} for Python programs requires only a few minutes to infer accurate grammars and produce thousands of valid inputs for formats such as \json.
Our approach is generic in its use of tools, as we could easily integrate different grammar learners or producers.  It is also versatile in its purposes, as the resulting grammars could also be used for activities such as input understanding, program understanding, parsing and translating inputs, or debugging.

The remainder of this paper is organized as follows.  \Cref{sec:example} illustrates our approach using arithmetic expressions as an example, devoting a section each to the individual steps from \Cref{fig:overview}:
\begin{enumerate}
\item \Cref{sec:testing} details how we generate inputs to systematically cover parsing alternatives.  
\item \Cref{sec:mining} shows how we use the resulting inputs and equivalence classes to induce high-quality grammars.
\item \Cref{sec:fuzzing} discusses how we use these grammars as producers, reentering grammar induction should generated inputs be rejected.
\end{enumerate}
\Cref{sec:evaluation} evaluates our publicly available \pyg prototype testing formats such as \ul and \json.  We find that \pyg achieves the same coverage as constraint-based alternatives; its inputs, however, are not only much more likely to be valid, they also cover and combine more features of the input language.  \Cref{sec:conclusion} closes with conclusion and future work.

\section{Our Approach in a Nutshell}
\label{sec:example}

To illustrate our approach, let us assume we want to exhaustively test some mystery program~$P$.  We know nothing about~$P$; in particular, we have no documentation or example inputs.  What we know, though, is that
\begin{enumerate*}
\item $P$ accepts some input~$I$ sequentially as a string of characters; and that
\item $P$ can tell us whether $I$ is a valid or an invalid input.
\end{enumerate*}
We further assume that we can \emph{observe}~$P$ processing~$I$: Specifically, we need to be able to observe the dynamic \emph{data flow of input characters} from~$I$ as $P$~processes them.

\subsection{Step 1: Testing a Parser}
\label{sec:testing}

In Step~1 (\Cref{fig:overview}), we explore the capabilities of $P$'s input parser by means of directed test generation.  The key idea is to observe all \emph{comparisons an input character goes through}, and systematically satisfy and cover alternatives, notably on rejection.

We start with an empty string as input, which is rejected as invalid
immediately as \emph{EOF} is encountered. The \emph{EOF} is detected
as any operation that tries to access past the end of given argument.
This error is fixed in the next round by testing $P$ with a random string,
say \<"A"{}> ($I = \<"A">$). Indeed, this input is also rejected by~$P$ as
invalid.
Before rejecting the input, though, $P$ checks $I$ for a number of properties: 
\begin{enumerate}
\item Does $I$ start with a digit?  \label{cond:digit}
\item Does $I$ start with a \<'('> character?  \label{cond:parenthesis}
\item Does $I$ start with \<'+'> or \<'-'>?  \label{cond:plusminus}
\end{enumerate}
Only after these checks fail does $P$ reject the input.

All these conditions are easy to satisfy, though---and this is a \emph{general} property of parsers, which typically only consider the single next character. 
Using a combination of depth-first and breadth-first search, our test generator picks one condition randomly. Satisfying \Cref{cond:digit}, it would produce a digit as input (say, \<"1"{}>).  This would now be accepted by~$P$ as valid, and we have generated our first input.

After the acceptance of \<"1"{}> as a partial input, $P$ conducts a check
to see if another character follows  \<"1"{}> by accessing the next character in
the input. Since $P$ reached the end of the string we consider the prefix as
valid and add another random character. This results in the new prefix \<"1B"{}>
which results in new conditions: Is the \<"B"{}> a digit? Or any of the
characters \<'+'>, \<'-'>, \<'*'>, or \<'/'>?  Again, one of these conditions
is chosen randomly, together with the prefix \<"1B"{}> seen so far.

In a consecutive execution with another random seed, the first condition to be addressed might be \Cref{cond:parenthesis}.  Satisfying this condition yields \<"("{}> which will again cause the parser reaching the end of the input, so we append a random character and get \<"(C"{}> as input. This is rejected, but only after again checking for a number of expected characters that could follow. These would be the same checks already performed on the input \<"A"{}>: digits, parentheses, \<'+'>, and \<'-'>. 
We randomly choose the condition \Cref{cond:plusminus}, where again the prefixes \<"(+"{}> and \<"(-"{}> would be invalid on their own, so we again choose one prefix for further computations.

By continuing this process, we thus obtain more and more inputs that systematically cover the capabilities of the parser.
In the end, we obtain a \emph{set of legal inputs that covers all the conditions encountered during parsing}:
\begin{alltt}
\hfill 1 \hfill 11 \hfill +1 \hfill -1 \hfill 1+1 \hfill 1-1 \hfill 1*1 \hfill 1/1 \hfill (1) \hfill
\end{alltt}
We see that our mystery program~$P$ in fact takes arithmetic expressions as inputs.

\subsection{Step 2: Inducing a Grammar}
\label{sec:mining}

In Step~2 (\Cref{fig:overview}), we take the generated inputs together with~$P$ to induce an \emph{input grammar}---that is, a context-free grammar which describes the input language of~$P$.  To this end, we feed the generated inputs into~$P$ while \emph{tracking their data flow,} notably into variables and function arguments.  

We find that an input such as \<"1+1"{}> flows into a function\linebreak \<parse\_expr()>, which
\begin{enumerate*}
\item first recursively invokes \<parse\_expr()> on the left \<"1"{}>,
\item then invokes \<parse\_binop()> on the \<"+"{}>, and 
\item finally recursively invokes \<parse\_expr()> on the right \<"1"{}>.
\end{enumerate*}
Tracking the recursive calls of \<parse\_expr()> on \<"1"{}>, we find that these invoke \<parse\_int()>, which in turn invokes \<parse\_digit()>, always passing the \<"1"{}> as argument.
From this sequence of calls, we can now induce a grammar rule, using these key ideas:
\begin{enumerate}
\item First, we can associate input fragments with the functions that successfully process them and assume that each input argument to a function represents a syntactic entity.  Hence, \<"1"{}> is a \emph{digit,} an \emph{integer,} and an \emph{expression;} \<"+"{}> is a \emph{binary operator;} and \<"1+1"{}> is an \emph{expression.}

\item Second, if some entity~$E$ is a substring of a larger entity~$E'$, we can derive a grammar rule decomposing~$E$ into~$E'$.  In the above case, we obtain rules such as
\begin{alltt}
\nonterm{Expr} \expandsto \nonterm{Int} | \nonterm{Expr} \nonterm{BinOp} \nonterm{Expr};
\nonterm{BinOp} \expandsto \term{+};
\nonterm{Int} \expandsto \nonterm{Digit}
\nonterm{Digit} \expandsto \term{1};
\end{alltt}

\item Third, during parser-directed test generation, we track \emph{equivalence classes} as induced by successful conditions.  We thus know that besides \<"1"{}>, any digit would have satisfied the conditions seen.  We can thus replace \<"1"{}> with the equivalence class of all digits:
\begin{alltt}
\nonterm{Digit} \expandsto \regex{[0-9]};
\end{alltt}
\todo{Say something about how to infer \nonterm{Int} \expandsto \nonterm{Digit}+?  -- AZ}

\item Finally, we can repeat the process for all inputs seen during the parser-directed test generation in Step~1.  This introduces \emph{alternatives} for all elements processed in the grammar, covering all operators and other syntactic features.  The resulting grammar (\Cref{fig:grammar}) represents all alternatives seen.

\begin{figure*}
\begin{alltt}
\nonterm{Expr} \expandsto \nonterm{Int} | \nonterm{UnOp} \nonterm{Expr} | \nonterm{Expr} \nonterm{BinOp} \nonterm{Expr} | \term{(} \nonterm{Expr} \term{)};
\nonterm{UnOp} \expandsto \term{+} | \term{-};
\nonterm{BinOp} \expandsto \term{+} | \term{-} | \term{*} | \term{/};
\nonterm{Int} \expandsto \nonterm{Digit}+
\nonterm{Digit} \expandsto \regex{[0-9]};
\end{alltt}
\vspace{-0.5\baselineskip}
\caption{Grammar induced from the inputs in Step~1}
\vspace{-0.5\baselineskip}
\label{fig:grammar}
\end{figure*}
\end{enumerate}
With this, we now have obtained a full description of $P$'s input language---%
without any sample inputs, specification, or model.

\subsection{Step 3: Grammar-Based Fuzzing}
\label{sec:fuzzing}

Grammars as obtained in Step~2 can serve many purposes.  We can use them to understand the structure of inputs, as well as the programs that process them.  We can use them to parse and process existing inputs, for instance to create detailed statistics on the occurrences of specific elements, or to protect programs against invalid inputs.  Our main application in this paper, though, is their use for \emph{test generation.}

Turning a grammar into a producer is a simple exercise.  Starting with the start symbol (\nonterm{Expr} in our case), we keep on replacing nonterminal symbols by one of the alternatives until only terminal symbols are left.  To avoid boundless expansion, we can set a limit on the maximum length of the string; once this is reached, we always prefer expansion paths that lead to terminal symbols.

This generation process now no longer requires any execution, instrumentation, or analysis of the program under test.  Hence, it is fast; and the strings generated can even be applied to some other program~$P'$ that shares the input language with~$P$.  A simple grammar producer can thus easily generate thousands to millions of inputs per minute, covering all kinds of symbols and their combinations.  This is what our technique produces: \emph{Given only a program~$P$, without any input samples, we obtain an input grammar that accurately describes the input language of~$P$, and consequently, can generate as many syntactically valid test inputs as desired.}

\begin{figure*}
\begin{alltt}\small
-++7 / +(9 - 6 / 7 + 5) - 1 + (0) / -75 * +(3 - 6 - 0 - 7)
9 + 4 - 3 + 7 / 7 + 3 / +3 * (9 - 2 * 9) - 8
++--+7 - (6 * 6 * 3) / (0 + 2) / +(5 / 6 / 5 + 3 * 1)
3 * 2764 + 1 / 0 * 4 / -5 / 6 * (1 * (8) + 9 / 4 * 0 * +4)
3 * 5 + 0 * 0 / 8 - 7 * 7 * ++(2 + 5 - 9 * 9)
+05834 * --(46 + +1 / +-+(-46 / 4) - --(63 - -(5 + 1 + +2 * 0 / 
    ++82 + (9 + 6)))) / -404471632
3 - 4 / 5 - 0 / 6 + 1 * 9 * 4 - +334
8 + 7 / 4 * 9 - (3 + 6 - (7)) + -0 - -+(5 + 8) * -++++5 - -2973
\end{alltt}
\vspace{-0.5\baselineskip}
\caption{Fuzzing output from the grammar in \Cref{fig:grammar}}
\vspace{-0.5\baselineskip}
\label{fig:fuzzing-output}
\end{figure*}

\begin{table*}
  \centering
  \renewcommand{\arraystretch}{1.3}
  \caption{Results of fuzzing with valid inputs of \pyg, \afl, and \klee.}
  \scalebox{0.85}{
  \begin{tabular}{@{}lrrrrrrrrrrrr@{}} \toprule
    Language & \multicolumn{1}{c}{Time} &   \multicolumn{3}{c}{\# Valid Inputs/\# Inputs} & \multicolumn{3}{c}{\SC} & \multicolumn{3}{c}{Maximal Input Length} \\
     & PYG (Step 1+2+3) & PYG & AFL & KLEE & PYG & AFL & KLEE  & PYG & AFL & KLEE  \\
    \midrule
    \ul & 66s + 53s + 28s & 782/1,000 & 0/36 & 463,165/1,028,735  & 55\%  & 0\% & 56\%  & 275  & 0 & 31   \\
    \mes & 96s + 1,7s + 18s  &  736/1,000  & 14/80 & 54,867/498,801   & 63\%  & 50\%  & 63\% & 200 &  23 & 16    \\    
    \json & 98s + 19s + 9s  &  778/1,000 & 19/39 & 12,575/41,625     & 43\%  & 23\%  & 43\%  & 81 & 29 & 31  \\
    \bottomrule
  \end{tabular} }
  \caption*{AFL and KLEE were given the same time as PYG for all subjects}
  \label{tbl:fuzz}
\end{table*}

\section{Initial Results}
\label{sec:evaluation}

We have implemented the above approach as a proof-of-concept prototype in Python, named \pyg.
We evaluate \pyg and all its parts on three different formats: \json{}~\cite{pyjson2018}, \ul{}\footnote{We manually converted the Java \ul parser \cite{pyurl2018} to Python.}, and \me{}~\cite{pymathexpr2018}. We used a coverage tool for Python~\cite{batchelder2018coverage} to compute the coverage the inputs achieve on the different subjects. For comparison, we used the \afl{}~\cite{AFLFuzz} random fuzzer and \klee{}~\cite{cadar2008klee}, a symbolic execution engine, both state of the art input generators. Since \klee is not available for Python, we generated inputs with \klee on a C parser of the respective input language and then executed the Python parser for this language with the generated inputs. \afl and \klee were both run with default settings.  \autoref{tbl:fuzz} summarizes our results, detailed in the remainder of this section.

\subsection{Execution Time}

We let \pdtg run until it produced 100~inputs.
The length of inputs produced by \pdtg is affected by the complexity of
input grammar.
In particular,
when considering nested grammars, each successive character might increase
the amount of nesting in the string produced, by adding a character---e.g.~`('---or close existing nested structures---e.g.~`)'. Since we are interested in valid strings, after a fixed number of characters is produced, we switch to a strategy designed to identify short suffixes that can complete the current string prefix.
The inputs from \pdtg  was used
to infer the grammar (Steps~1 and~2); we then used this grammar to produce 1,000~inputs (Step~3). For producing samples from the grammar, we chose to limit the number of symbols expanded to~100 before applying heuristics to complete the string generation.

\autoref{tbl:fuzz} reports the \pyg execution times broken down per step; Steps 1~and~2 need to be run once per program, Step~3 for every 1,000 inputs generated.  Note that switching from Python to C would speed up all three steps further, especially Step 3.

For comparison, we let \afl and \klee run as long as all three phases of \pyg and assessed the resulting test cases.  \afl has no built-in limit to how long it will run and produce inputs; \klee stops as it has explored all paths, but would not reach this limit within the execution time of \pyg.

\subsection{Input Validity}

For all three subjects, between 73\%~and~78\% of all inputs generated by \pyg would be valid; the remainder is invalid due to overgeneralization in Step~2.  For \afl, we only report those inputs where it found a new path (which is the default setting); only between 0\% and 50\% of these inputs, though, are valid.  \klee produced thousands to millions of inputs, with 25\% to 46\% being valid.  Most of the inputs of \afl and \klee exercise handling of syntax errors.\footnote{For \ul, actually \emph{none} of the inputs generated by \klee would be valid in the original Python subject because the C~subject we applied \klee on would erroneously accept URLs without a protocol prefix.  For fairness, we therefore changed the Python parser to also accept URLs without prefix.}

\subsection{Coverage}

Let us now come to the one metric typically used to compare the performance of test generators---coverage.  We only report coverage of code handling valid inputs, as this would be the code that actually holds program functionality.  (As discussed before, if one wanted to deliberately produce invalid inputs, \afl would probably be the best choice.)  \pyg and \klee achieve a very similar coverage.  The only 1-point difference is in \ul, where \klee explores \ul queries (prefixed by \<'?'>) and \pyg doesn't; the reason is that (a) the \ul parser accepts any string after the hostname, with no special provisions for \<'?'> (queries) or \<'\#'> (anchors); and (b) \pyg's grammar inference does not generalize the characters to include \<'?'> characters.  Apart from \<'?'>, in all three cases, the coverage achieved by \pyg is the maximum one can achieve on these subjects using valid inputs.

\subsection{Input Quality}

A good test case will not only cover code, but also explore \emph{combinations} of features to thoroughly test their possible interactions and interference.  As a very simple assessment of how our inputs fare in this regard, we take a look at the generated valid inputs with \emph{maximum length}, for example for \json:  

\begin{itemize}
  \item The longest \pyg input covers and combines \json elements such as arrays, objects, strings, and numbers\footnote{It also exercises a bug in the \microjsonpy parser.}:
\begin{alltt}
[false ,[\{  "o":\{    ,   "$dYPrlj@?BR":
 [+ ]"S|+|4GzCW(C":-94\}\}  ],
 [false,null]]
\end{alltt}
\item The longest \afl input consists of 29~periods (\<'.'>)
\item The longest \klee input consists of the keyword \texttt{null}, followed by 27~8-bit ASCII~255 (\<'\"y'>) characters.
\end{itemize}
We see that the \pyg input is \emph{considerably richer in syntax and semantics.}  For \mes, the situation is the same: \afl and \klee produce a long single number, whereas \pyg combines elements as in \Cref{fig:fuzzing-output}; only for \ul does the longest \klee input actually cover elements of the URL structure.

One may argue that \pyg is set to produce 100~symbols and thus longer inputs than \klee with 30 characters per input\footnote{\afl can generate inputs of arbitrary size.}. But then, the search effort reduces for \klee if the input size is small while still getting the chance to produce complex inputs. But even with 30 characters \klee is not able to produce any complex inputs that make use of the size.
Furthermore, and this is precisely the point: When producing from a grammar, not only do we get well-structured complex inputs, as with \pyg.  \emph{For a tester, it also is very easy to control input length or depth, to emphasize or de-emphasize symbols, or to favor or cover specific combinations of symbols.}  This is only possible with well-structured and well-readable grammars, whose inference thus contributes to the quality of test cases and the potential of our approach.

\section{Conclusion and Consequences}
\label{sec:conclusion}

We have shown that it is possible to \emph{determine the input language from a given program alone, without requiring input samples.}  This finding has many applications throughout programming languages and software engineering, for instance in understanding both input and program structure.  First and foremost, though, this is an important step forward for test generation at the system level, which so far required either 
\begin{enumerate*}
\item a model for the input (say, an input grammar or a state model), or
\item a set of sample inputs (which would be mutated, evolved, or abstracted into a model).
\end{enumerate*}
In contrast, our approach makes it possible to take a given program only, infer its input language automatically, and immediately use this for producing syntactically valid inputs with high coverage---all without any human effort, as we demonstrate in this work.  At the same time, the inferred grammars give testers (and tools) control over which and how many elements should be covered and generated, targeting features and feature combinations much better than mutation-based or constraint-based approaches.

While Python is a great language for prototyping, most common input formats are parsed in C programs and libraries; therefore, we are currently implementing our approach for C programs.  In the meantime, a replication package with \pyg source code as well as all experimental settings and data is available~\cite{pygmaliondata}.

\printbibliography
\end{multicols}
\end{document}